\documentclass[aps,prd,preprint,showpacs,nofootinbib,11pt]{revtex4-1}
\usepackage{latexsym,bm,amsmath,amssymb}

\DeclareMathAlphabet{\mathpzc}{OT1}{pzc}{m}{it}

\begin{document}

\title{On the thermodynamics of Lifshitz black holes} 

\author{Deniz Olgu Devecio\u{g}lu}
\email{dedeveci@metu.edu.tr}
\author{{\"O}zg{\"u}r Sar{\i}o\u{g}lu}
\email{sarioglu@metu.edu.tr}
\affiliation{Department of Physics, Faculty of Arts and  Sciences,\\
             Middle East Technical University, 06800, Ankara, Turkey}

\date{\today}

\begin{abstract}
We apply the recently extended conserved Killing charge definition of Abbott-Deser-Tekin formalism
to compute, for the first time, the energies of analytic Lifshitz black holes in higher dimensions. We then
calculate the temperature and the entropy of this large family of solutions, and study and discuss the first 
law of black hole thermodynamics. Along the way we also identify the possible critical points of the
relevant quadratic curvature gravity theories. Separately, we also apply the generalized Killing charge 
definition to compute the energy and the angular momentum of the warped AdS$_3$ black hole solution 
of the three-dimensional New Massive Gravity theory.
\end{abstract}

\pacs{04.50.Gh, 04.20.Cv, 04.70.-s}

\maketitle

\section{\label{intro} Introduction}

In a recent paper \cite{biz}, we have extended the conserved Killing charge definition of the Abbott-Deser-Tekin
(ADT) \cite{des1,des2,des3} procedure valid for asymptotically constant-curvature spacetimes to a more general form,
which enables it to work also for geometries that asymptote to arbitrary backgrounds with at least one Killing isometry
and that solve quadratic curvature gravity models in generic $D$ dimensions. We have also shown in \cite{biz} that
this generalization is background gauge invariant and correctly reduces to its counterpart when the background is a 
space of constant curvature.

The motivation behind \cite{biz} was to a great extend coming from the challenge to compute the energy of a large 
family of analytic Lifshitz black holes \cite{giri2} that are solutions of quadratic curvature gravity theories in 
various dimensions. The Lifshitz spacetimes, which these black hole solutions asymptote to, have the same 
symmetries as certain models of condensed matter systems (see \cite{giri2,giri1,bala1} and the references therein 
for details). The first analytic Lifshitz black hole, which however is not a member of the set of black holes 
we consider here, was given in \cite{bala2}. There are also a number of such black hole solutions that were obtained
numerically (see e.g. the references in \cite{bala1,bala2}), but these are not suitable for the formalism we employ
here. The first member of the set of Lifshitz black holes that we study in what follows was the four-dimensional 
solution given in \cite{cinli}. However the large family presented in \cite{giri2} is much more general and contains 
the former as a particular point. The authors of \cite{giri2} have also found the three-dimensional Lifshitz black hole 
\cite{giri1}, which is a solution of the New Massive Gravity (NMG) theory \cite{hohm} of late. 

These Lifshitz black holes and their asymptotes are not spaces of constant curvature and globally possess a 
timelike Killing vector, and thus provide perfect examples which our generalized formulation can be applied to. 
Hence the main purpose of this work is to employ our generalized charge definition to give an answer to the 
open problem of computing the energies of these black hole solutions. We want to emphasize that this in itself 
is quite a nontrivial task and, to our knowledge, has never been undertaken for the entire family of Lifshitz
black holes given in \cite{giri2}.

The next issue we deal with concerns the computation of the temperature and the entropy of these black holes.
Once all the necessary ingredients have been obtained, one is in a position to bring all these together in the
first law of black hole thermodynamics, which surprisingly turns out not to follow automatically for all of these
solutions. We may interpret this as an indication that not all of these black holes are thermodynamically stable. 
Yet we also discuss other possible reasons as to why this is so. As a by product of our energy calculations, we
also identify the possible critical points, in the sense of \cite{dllpst}, of the theories under consideration.

The final task we undertake in this work is somewhat separate from the overall consideration so far. 
We want to use this opportunity to also compute the energy and the angular momentum of the warped 
AdS$_3$ black hole solution \cite{clem} of the NMG theory as a final quest. These conserved charges 
have already been calculated by other methods previously, but this provides another important test ground 
for our generalized charge expression, all the more so since, to our knowledge, the warped AdS$_3$ black 
hole spacetime is the only stationary geometry which involves rotation and asymptotes to the so called warped 
AdS$_3$ spacetime, which definitely is not a space of constant curvature, and, thus, is perfect for our formulation.

The organization of the paper is as follows: In section \ref{charge}, we give a brief summary of the ADT
formulation and write down the generalized Killing charge found in \cite{biz} for quadratic curvature gravity 
theories. Section \ref{enerLBH} starts by presenting a general expression for the energy of the Lifshitz black
holes, and continues with an analogous treatment that yields the entropy of these spacetimes. The subsections
of section \ref{enerLBH} are devoted to the computation of these physical quantities, together with the 
temperature, of various subfamilies of Lifshitz black holes and to a discussion of the first law of
black hole thermodynamics. Separate from the earlier exposition, section \ref{enerNMG} basically deals with 
the computation of the conserved charges of the warped AdS$_3$ black hole solution. We then finish up with 
our conclusions.

\section{\label{charge} The generalized Killing charge for quadratic curvature gravity theories}

Let us begin by briefly summarizing the ADT procedure and display the general charge expression obtained in
\cite{biz} for quadratic curvature gravity theories. 

One starts with a local gravity action with field equations \( \Phi_{ab} = \kappa \, \tau_{ab}, \) that admits a
background metric $\bar{g}_{ab}$ as a solution to the source-free theory such that \( \bar{\Phi}_{ab}(\bar{g}) = 0 \) 
for \( \tau_{ab}= 0 \). One then splits the metric as \( g_{ab} = \bar{g}_{ab} + h_{ab}, \) where the deviation
$h_{ab}$ is assumed to vanish sufficiently rapidly as one asymptotically approaches to the boundary of spacetime
whose metric is given far away from the sources by the background metric $\bar{g}_{ab}$. The full field equations
are then linearized about the background so that \( \Phi_{ab}^{L} = \kappa \, T_{ab}, \) where $\Phi_{ab}^{L}$
is linear in $h_{ab}$, and $T_{ab}$ (already containing $\tau_{ab}$) acts as a total source. Provided that the
background metric $\bar{g}_{ab}$ admits at least one Killing isometry generated by a Killing vector $\bar{\xi}^{a}$, 
with \( \bar{\nabla}_{a} \bar{\xi}_{b} +  \bar{\nabla}_{b} \bar{\xi}_{a} = 0 \), and using the fact that 
$\Phi_{ab}^{L}$ is conserved by linearized diffeomorphism invariance, one is then able to obtain
\[ \bar{\nabla}_{a} (T^{ab} \, \bar{\xi}_{b}) = 0 = \partial_{a} (\sqrt{-\bar{g}} \, T^{ab} \, \bar{\xi}_{b}), \]
which can be used to arrive at the required charge definition.

The action of a generic quadratic curvature gravity theory in $D$-dimensions reads
\begin{equation}
 I = \int \, d^{D}x \, \sqrt{-g} \, {\mathcal L} \equiv \int \, d^{D}x \, \sqrt{-g} \, \Big( \frac{1}{\kappa} (R + 2 \Lambda_{0}) 
+ \alpha R^{2} + \beta R_{ab} R^{ab} + \gamma (R_{abcd} R^{abcd} - 4 R_{ab} R^{ab} + R^2) \Big), \label{act}
\end{equation}
where $\Lambda_{0}$ is the bare cosmological constant and $\kappa$ is related to the $D$-dimensional
Newton's constant. Following the steps outlined above, one finds, as explained in \cite{biz}, that
\begin{equation}
 \Phi^{ab}_{L} \bar{\xi}_{b} = \bar{\nabla}_{b} {\mathcal F}^{ab} + h^{ab} \bar{\Phi}_{bc} \bar{\xi}^{c} 
+  \frac{1}{2} \bar{\xi}^{a} \bar{\Phi}_{bc} h^{bc} - \frac{1}{2} h \bar{\Phi}^{ab} \bar{\xi}_{b} \,, \label{phixi}
\end{equation}
where
\begin{equation}
 {\mathcal F}^{ab} = - {\mathcal F}^{ba} =  \frac{1}{\kappa} {\mathcal F}^{ab}_{E} + \alpha {\mathcal F}^{ab}_{\alpha} 
 + \beta {\mathcal F}^{ab}_{\beta} + \gamma {\mathcal F}^{ab}_{\gamma} \,, \label{fab}
\end{equation}
with
\begin{eqnarray}
 {\mathcal F}^{ab}_{E} & \equiv & \bar{\xi}_{c} \bar{\nabla}^{[a} h^{b]c} + \bar{\xi}^{[b} \bar{\nabla}_{c} h^{a]c} 
 + h^{c[b} \bar{\nabla}_{c} \bar{\xi}^{a]} + \bar{\xi}^{[a} \bar{\nabla}^{b]} h + \frac{1}{2} h \bar{\nabla}^{[a} \bar{\xi}^{b]} , 
 \label{einchar} \\
 {\mathcal F}^{ab}_{\alpha} & \equiv & 2 \bar{R} {\mathcal F}^{ab}_{E} + 2 \bar{\xi}^{[b} h^{a]c} \bar{\nabla}_{c} \bar{R}
 + 4 \bar{\xi}^{[a} \bar{\nabla}^{b]} R_{L} + 2 R_{L} \bar{\nabla}^{[a} \bar{\xi}^{b]} , \label{alchar} \\
 {\mathcal F}^{ab}_{\beta} & \equiv & \bar{\xi}^{[a} \bar{\nabla}^{b]} R_{L} + 2 \bar{\xi}^{c} \bar{\nabla}^{[b} (R^{a]}\,_{c})_{L} 
 + 2 (R^{[b}\,_{c})_{L} \bar{\nabla}^{a]} \bar{\xi}^{c} + h_{cd} \bar{\xi}^{[b} \bar{\nabla}^{a]} \bar{R}^{cd} 
 + 2 h^{c[a} \bar{\xi}_{d} \bar{\nabla}_{c} \bar{R}^{b]d} + 2 \bar{R}^{c[a} \bar{\xi}_{d} \bar{\nabla}_{c} {h^{b]d}} \nonumber \\
 & & + h \bar{\xi}_{c} \bar{\nabla}^{[b} \bar{R}^{a]c} + 2 \bar{R}^{c[b} h^{a]d} \bar{\nabla}_{c} \bar{\xi}_{d} 
 + h \bar{R}^{c[b} \bar{\nabla}^{a]} \bar{\xi}_{c} + 2 \bar{R}^{c[b} \bar{\xi}^{d} \bar{\nabla}_{d} {h^{a]}\,_{c}} 
 + \bar{\xi}^{[a} \bar{R}^{b]c} \bar{\nabla}_{c} h + \bar{R}^{cd} \bar{\xi}^{[a} \bar{\nabla}^{b]} h_{cd} \nonumber \\
 & & + 2 \bar{\xi}^{d} \bar{R}_{cd} \bar{\nabla}^{[b} h^{a]c} + 2 \bar{R}^{c[a} \bar{\xi}^{b]} \bar{\nabla}^{d} h_{cd} 
 + 2 \bar{\xi}^{d} \bar{R}^{c[b} \bar{\nabla}^{a]} h_{cd} , \label{bechar} \\
 {\mathcal F}^{ab}_{\gamma} & \equiv & 2 \bar{R} {\mathcal F}^{ab}_{E} + 2 \bar{R}^{[ba]cd} \bar{\xi}_{d} \bar{\nabla}_{c} h
 + 4 \bar{\xi}_{c} \bar{R}^{c[b} \bar{\nabla}^{a]} h + 4 \bar{R}^{c[a} \bar{\xi}^{b]} \bar{\nabla}_{c} h 
 + 2 h \bar{R}^{c[ab]d} \bar{\nabla}_{d} \bar{\xi}_{c} + 4 h \bar{R}^{c[a} \bar{\nabla}^{b]} \bar{\xi}_{c} \nonumber \\
 & & + 4 \bar{\xi}_{d} \bar{R}^{dec[a} \bar{\nabla}_{c} h^{b]}\,_{e} + 4 \bar{\xi}_{d} \bar{R}^{dec[a} \bar{\nabla}_{e} h^{b]}\,_{c} 
 + 4 \bar{\xi}_{d} \bar{R}^{dec[b} \bar{\nabla}^{a]} h_{ce} + 4 \bar{\xi}^{[a} \bar{R}^{b]cde} \bar{\nabla}_{d} h_{ce} 
 + 2 \bar{R}^{[ab]cd} \bar{\xi}^{e} \bar{\nabla}_{c} h_{de} \nonumber \\
 & & + 2 \bar{\xi}_{d} h_{ce} \bar{\nabla}^{c} \bar{R}^{[ab]de} + 4 \bar{\xi}_{d} \bar{R}^{d[a} \bar{\nabla}_{c} h^{b]c} 
 + 4 h_{ce} \bar{R}^{dec[a} \bar{\nabla}^{b]} \bar{\xi}_{d} + 4 \bar{R}^{[ab]cd} h_{ce} \bar{\nabla}_{d} \bar{\xi}^{e} 
 + 4 \bar{R}_{cd} \bar{\xi}^{d} \bar{\nabla}^{[b} h^{a]c} \nonumber \\
 & & + 4 \bar{\xi}^{d} \bar{R}^{c[a} \bar{\nabla}^{b]} h_{cd} + 4 \bar{\xi}^{d} \bar{R}_{c}\,^{[b} \bar{\nabla}_{d} h^{a]c}
 + 4 \bar{\xi}^{d} \bar{R}^{c[b} \bar{\nabla}_{c} h^{a]}\,_{d} + 8 \bar{R}^{cd} \bar{\xi}^{[a} \bar{\nabla}_{c} h^{b]}\,_{d} 
 + 4 \bar{R}^{cd} \bar{\xi}^{[b} \bar{\nabla}^{a]} h_{cd} \nonumber \\
 & & + 2 \bar{R}^{cd} h_{cd} \bar{\nabla}^{[b} \bar{\xi}^{a]} + 4 \bar{\xi}_{d} h^{c[a} \bar{\nabla}_{c} \bar{R}^{b]d} 
 + 4 h^{cd} \bar{\xi}^{[b} \bar{\nabla}_{c} \bar{R}^{a]}\,_{d} + 4 h^{c[a} \bar{R}^{b]d} \bar{\nabla}_{c} \bar{\xi}_{d} 
 + 8 h_{cd} \bar{R}^{c[b} \bar{\nabla}^{a]} \bar{\xi}^{d} \nonumber \\
 & & + 2 \bar{\xi}^{[a} h^{b]c} \bar{\nabla}_{c} \bar{R} , \label{gachar}
\end{eqnarray}
and one should recall that the background is to be chosen as \( \bar{\Phi}_{ab}(\bar{g}) = 0 \). (See \cite{biz} for details 
and the conventions used here.)

Hence the relevant charge expression is given by 
\begin{equation}
 Q(\bar{\xi}) = \int_{\Sigma} d^{D-1}x \, \sqrt{|\sigma|} \, n_{a} \, \Phi^{ab}_{L} \, \bar{\xi}_{b} =
 \int_{\partial \Sigma} d^{D-2}x \, \sqrt{|\sigma^{(\partial \Sigma)}|} \, n_{a} \, s_{b} \, {\mathcal F}^{ab} \,, 
\label{genchar}
\end{equation}
where the Stokes's theorem is used to relate the integral over the ($D-1$)-dimensional hypersurface $\Sigma$ (with its
induced metric $\sigma$ and its unit normal $n^{a}$) to its integral over the ($D-2$)-dimensional boundary
$\partial \Sigma$ (with its induced metric $\sigma^{(\partial \Sigma)}$ and its unit normal $s^{b}$). 

\section{\label{enerLBH} The energy and the entropy of Lifshitz black holes}

We first apply the general charge formula (\ref{genchar}) reviewed in the previous section to compute
the energies of Lifshitz black holes presented in \cite{giri2}. We then employ Wald's formula \cite{wald}
to also find the entropy of the Lifshitz black holes. At first we keep to a generic case and later specialize
to specific examples in the subsequent subsections.

Lifshitz black hole solutions are of the generic form
\begin{equation}
 ds^2 = - \frac{r^{2 z}}{\ell^{2 z}} f(r) dt^2 + \frac{\ell^2}{r^2} \, \frac{dr^2}{f(r)} + \frac{r^2}{\ell^2} d\vec{x}^2, 
 \label{lifbh}
\end{equation}
where the function $f(r)$  and the range of the dynamical exponent $z$ depend on the specific quadratic 
curvature gravity theory under consideration, and $\vec{x}$ is a $(D-2)$-dimensional vector denoting the 
`angular variables' of the $D$-dimensional spacetime. Typically the function $f(r)$ has a behavior such 
that the boundary of the Lifshitz black hole is located at infinite $r$, and the background is easily seen 
to be given by the Lifshitz spacetime, whose metric is obtained from (\ref{lifbh}) by taking $f(r) \to 1$ 
as $r \to \infty$, and reads
\begin{equation}
 ds^2 = - \frac{r^{2 z}}{\ell^{2 z}} dt^2 + \frac{\ell^2}{r^2} \, dr^2 + \frac{r^2}{\ell^2} d\vec{x}^2. 
 \label{lifsp}
\end{equation}
Note that the background (\ref{lifsp}) obviously admits \( \bar{\xi}^{a} = - (\partial/\partial t)^{a} \) as a timelike Killing
vector and the most relevant conserved quantity that can be computed using the formula (\ref{genchar}) is 
the energy of Lifshitz black holes. Thus one is naturally led to choose $\Sigma$ in (\ref{genchar}) to be 
a hypersurface of constant time, which is what we do in what follows. Taking into account the form of the 
background, one then also has to set \( n^{a} = - (\ell^{z}/r^{z} ) \delta^{a}_{t} \) and 
\( s^{a} = (r/\ell) \delta^{a}_{r} \), so that these are the required unit vectors that satisfy the properties 
mentioned above. Putting all of these together, one is thus led to
\begin{equation}
 E = \int_{\partial \Sigma} d^{D-2}x \, \sqrt{|\sigma^{(\partial \Sigma)}|} \, n_{t} \, s_{r} \, {\mathcal F}^{tr}
 = \lim_{r \to \infty} \int_{\partial \Sigma} d^{D-2}x \, \frac{r^{z+D-3}}{\ell^{z+D-3}} \, {\mathcal F}^{tr} 
 = \Omega_{D-2} \, \Big( \lim_{r \to \infty} \frac{r^{z+D-3}}{\ell^{z+D-3}} \, {\mathcal F}^{tr} \Big)
\label{ener}
\end{equation}
for the energy $E$ of the Lifshitz black holes. Here $\Omega_{D-2}$ represents the finite contribution of the 
$(D-2)$-dimensional integration over the ranges of the angular variables $\vec{x}$, and 
\( {\mathcal F}^{tr} = {\mathcal F}^{tr}(r) \) is to be calculated using (\ref{fab}) to (\ref{gachar}) above.

Meanwhile, the entropy of the Lifshitz black holes can be calculated using Wald's formula \cite{wald}
which in our conventions reads
\begin{equation}
 S_{W} = - 2 \pi \oint_{\mathpzc{H}} d^{D-2}x \, \sqrt{|\mathpzc{h}_{+}|} \, \frac{\delta {\mathcal L}}{\delta R_{abcd}} \,
  \varepsilon_{ab} \,  \varepsilon_{cd} \,, \label{w1} 
\end{equation}
where the integral is to be taken at the spatial section $\mathpzc{H}$ of the event horizon, which is located at 
$r = r_{+}$ that is clearly found by the largest positive real root of $f(r)$, i.e. $f(r_{+}) = 0$. Here $|\mathpzc{h}_{+}|$
denotes the determinant of the induced metric on $\mathpzc{H}$, whereas $\varepsilon_{ab}$ is the binormal to
the event horizon and one is to employ the Lagrangian ${\mathcal L}$ given in (\ref{act}) above.

For the Lifshitz black holes, whose metric is of the form (\ref{lifbh}), the binormal $\varepsilon_{ab}$ follows from the 
timelike Killing vector  \( \chi^{a} = - (\partial/\partial t)^{a} \), which becomes null at the event horizon, and is given by
\[ \varepsilon_{ab} = - \varepsilon_{ba} \equiv \frac{1}{\varkappa} \nabla_{a} \chi_{b} = 
 \Big( \frac{r^{2z-1}}{\ell^{z-1}} \frac{(2 z f(r) + r f^{\prime}(r))}{r_{+}^{z+1} f^{\prime}(r_{+})} \Big) 
 \delta_{ar} \delta_{bt} \,, \]
where prime indicates derivative with respect to the argument and the surface gravity $\varkappa$ is
given by
\[ \varkappa \equiv \sqrt{- \frac{1}{2} (\nabla_{a} \chi_{b}) (\nabla^{a} \chi^{b}) } 
 = \frac{1}{2} \frac{r_{+}^{z+1}}{\ell^{z+1}} f^{\prime}(r_{+}) \,. \]
Meanwhile, it follows easily from the generic Lagrangian ${\mathcal L}$ in (\ref{act}) that
\[  \frac{\delta {\mathcal L}}{\delta R_{abcd}} \, \varepsilon_{ab} \,  \varepsilon_{cd} = 
 \Big( \frac{1}{\kappa} + 2 (\alpha + \gamma) R \Big) \varepsilon_{ab} \, \varepsilon^{ab} 
 + 2 (\beta - 4 \gamma) R_{bc} \, \varepsilon^{ab} \, \varepsilon_{a}\,^{c} 
 + 2 \gamma R_{abcd} \, \varepsilon^{ab}  \, \varepsilon^{cd} \,. \]

Putting all of these ingredients together leads to
\begin{equation}
 S_{W} = - 2 \pi \, \Omega_{D-2} \, \frac{r_{+}^{D-2}}{\ell^{D-2}} \, 
 \Big[ \frac{\delta {\mathcal L}}{\delta R_{abcd}} \, \varepsilon_{ab} \,  \varepsilon_{cd} \Big]_{r = r_{+}} \, \label{w2}
\end{equation}
for the Wald entropy $S_{W}$ of the Lifshitz black holes, whereas the temperature $T$, as usual, can be taken as
\( T = \varkappa / (2 \pi) \).

In the remainder of this section, we use (\ref{ener}) to compute the energies and (\ref{w2}) to find the entropies 
of the Lifshitz black holes presented in \cite{giri2}. To keep the discussion concise, we will, whenever we can, 
refrain from duplicating the formulas already given in \cite{giri2} and refer the reader to that reference for relevant 
details.

\subsection{\label{sec2} The energy and the entropy of $R^2$-corrected Lifshitz black holes}

This class of solutions was presented in section 2 of \cite{giri2} and their most general form is given by (\ref{lifbh}),
where 
\begin{equation}
 f(r) = 1 - \frac{M^{-} \ell^{p_{-}}}{r^{p_{-}}} + \frac{M^{+} \ell^{p_{+}}}{r^{p_{+}}} \quad \mbox{and} \quad
 p_{\pm} = \frac{3 z +2 (D-2) \pm \sqrt{z^2 + 4(D-2)(z-1)}}{2} \,,
\label{fofsec2}
\end{equation}
with $M^{\pm}$ as the two free parameters. The spacetime (\ref{lifbh}) (with (\ref{fofsec2})) solves the 
quadratic curvature gravity theory (\ref{act}) when
\[ \kappa = 1 \,, \quad \Lambda_{0} = \frac{2 z^{2} + (D-2)(2 z + D-1)}{4 \ell^{2}} \,, \quad 
 \alpha = \frac{1}{8 \Lambda_{0}} \,, \quad \beta = \gamma = 0 \]
in our conventions. However, it turns out that the curvature scalar of this spacetime is exactly 
\( R = - 4 \Lambda_{0} \), and, as discussed in \cite{giri2}, the gravity action can be put into the form
\[  I = \frac{1}{8 \Lambda_{0}} \int \, d^{D}x \, \sqrt{-g} \, (R + 4 \Lambda_{0})^{2} \,, \]
which renders it impossible for this theory to have a scalar-tensor cousin that is obtained by employing 
conformal transformations. Finally, it should be noted that the spacetime (\ref{lifbh}) (with (\ref{fofsec2})) 
represents a black hole if $p_{\pm} > 0$, which happens when \( z \geq z_{+} = 4 - 2 D + 2 \sqrt{(D-1)(D-2)} \).

As an illustrative example of how the calculation goes in the general case, we now consider the Lifshitz black hole 
given in \cite{cinli}, which is just a specific member of this class with $D = 4$, $z = 3/2$, $M^{+} = 0$ and 
$\Lambda_{0} = 33/(8 \ell^{2})$. In this case, one finds that
\[ {\mathcal F}^{tr}_{E} = \frac{\ell M^{-} r}{r^{3} - \ell^{3} M^{-}} \quad \mbox{and} \quad
   {\mathcal F}^{tr}_{\alpha} = \frac{3 M^{-} (22 r^9 + \ell^{3} M^{-} r^6 + 97 \ell^{6} (M^{-})^2 r^3 - 36 \ell^{9} (M^{-})^3)}
   {2 \ell r^2 (\ell^{3} M^{-} - r^{3})^3} \,, \]
which leads to
\[ E = \Omega_{2} \, \Big( \lim_{r \to \infty} \frac{r^{5/2}}{\ell^{5/2}} \, {\mathcal F}^{tr} \Big) =
  \Omega_{2} \, \Big( \lim_{r \to \infty} \frac{3 \ell^{3/2}  (M^{-})^2 r^{1/2}}{22 (\ell^{3} M^{-} - r^{3})^3} 
 (15 r^6 + 25 \ell^{3} M^{-} r^{3} -12 \ell^{6} (M^{-})^2) \Big) = 0 \,. \]
It must be noted that the vanishing of energy was deduced, although indirectly, also in \cite{cinli} for the special solution
just considered.

It turns out that this unusual result remains intact, i.e. $E = 0$, for the most general Lifshitz
black hole (\ref{lifbh}) (with (\ref{fofsec2})) in generic $D$ dimensions. Moreover, the other siblings of this
solution, which are given by Eqs. (2.7), (2.9), (2.11) and (2.12) in \cite{giri2} (these are obtained by setting a
specific relation between the parameters $M^{\pm}$ to arrive at the extremal black hole, taking $z = z_{+}$, 
finding the extremal version of the latter and setting $z = 1$, respectively), all have $E = 0$ in generic $D$ 
dimensions. We have verified this result by meticulously working out inequalities implied by the condition 
$z \geq z_{+}$ in rather cumbersome expressions obtained from (\ref{ener}) and checked it with a computer code. 

However, this result is not too surprising after all. Considering the remark on the gravitational action above,
it is straightforward to observe that the action attains its minimum null value at the solution (\ref{lifbh}) 
(with (\ref{fofsec2})) \cite{giri2}. In fact, since \( 2 \alpha R + 1/\kappa = 0 \) for this class of solutions, 
it is fairly straightforward to show, using Wald's formula (\ref{w2}), that the entropy of this spacetime 
vanishes identically, i.e. $S_{W} = 0$, as well. We believe this lends more support on the argument that 
this spacetime can be thought of as a gravitational instanton \cite{giri2}.

\subsection{\label{sec31} The energy and the entropy of Lifshitz black holes for $z>2-D$}

This class of solutions is in section 3.1 of \cite{giri2} and they are in the generic form (\ref{lifbh}), but now
\begin{equation}
 f(r) = 1 - \frac{M \ell^{(z+D-2)/2}}{r^{(z+D-2)/2}} \,,
\label{fofsec31}
\end{equation}
where $M$ is a free parameter. The spacetime (\ref{lifbh}) (with (\ref{fofsec31})) solves the 
quadratic curvature gravity theory (\ref{act}) for $D \geq 5$, where
\begin{equation}
 \kappa = 1 \,, \quad \Lambda_{0} = - \lambda \,, \quad \gamma = \beta_{3} \,, \quad
 \alpha = \beta_{1} - \beta_{3} \,, \quad \beta = \beta_{2} + 4 \beta_{3} \,, 
\label{param}
\end{equation}
in terms of the parameters $\lambda$, $\beta_{1}$, $\beta_{2}$ and $\beta_{3}$ (that are
given by Eqs. (3.4b)-(3.4e) of \cite{giri2}) in our conventions. It should also be noted that the 
solution (\ref{lifbh}) (with (\ref{fofsec31})) represents a Lifshitz black hole when $z>2-D$.

The calculation of energy follows in the lines of the computation done in the previous
subsection, but the steps involved are much more complicated this time. Rather than giving the 
details, we prefer to simply state the final answer. For generic $D \geq 5$ and $z>2-D$, the 
energy is found as
\begin{equation} 
 E = \frac{M^2 \, (z-1) \, p(z)}{\ell \, (z+D-2) \, q(z)} \, \Omega_{D-2} \,, \label{ener31}
\end{equation}
where
\begin{eqnarray*}
 p(z) & = & - 8 (D-2) \Big( 9 z^4 - (D+5) z^3 - (D-2) \big[ 3 (D-5) z^2 + (D^2 - 5 D + 10) z - (D^2-4) \big] \Big) \,, \\
 q(z) & = & 27 z^4 - 4 (27 D - 45) z^3 - (D-2) \big[ 2 (5 D - 116) z^2 + 4 (D^2 - D +30) z + (D+2) (D-2)^2 \big] \,.
\end{eqnarray*}
Similarly, rather than giving the technical details, we again write down the temperature and the Wald entropy for the 
generic case. They are
\begin{equation}
 T = \frac{(z+D-2)}{8 \pi \ell} \, M^{\frac{2 z}{z+D-2}} \quad \mbox{and} \quad
 S_{W} = - 4 \pi \, \Omega_{D-2} \, \frac{(3 z^2 + (D^2 - 4)) \, (D^2 - (2-3z)^2)}{q(z)} \, M^{\frac{2(D-2)}{z+D-2}} \,,
 \label{ent31}
\end{equation}
and note that $T>0$ for generic values of $z>2-D$.

There are various remarks that need to be made at this point. First of all, the conformal limit $z = 1$ of this 
family (\ref{lifbh}) (with (\ref{fofsec31})) is an asymptotically AdS black hole solution \cite{giri2} of (\ref{act}) with 
\begin{equation}
 \kappa = 1 \,, \quad \Lambda_{0} = \frac{(D-1)(D-2)}{4 \ell^2} \,, \quad \gamma = \frac{\ell^2}{2 (D-3)(D-4)} \,, \quad
 \alpha = \beta = 0 \,. 
 \label{paramz=1}
\end{equation}
Perhaps unexpectedly though, it immediately follows from (\ref{ener31}) that the conformal limit $z = 1$ of this family has 
$E = 0$ for all $D \geq 5$. This may seem a bit counterintuitive but it turns out to be automatic, since setting $z = 1$ at 
the very beginning indeed leads nontrivially to $\Phi^{ab}_{L} = 0$ for all $D \geq 5$. In fact, since $E = 0$, the 
conformal limit $z = 1$ of this family may very well describe the critical points \cite{dllpst} of the relevant 
Einstein-Gauss-Bonnet theory in $D \geq 5$. However, one finds in this case that the temperature and the Wald 
entropy read
\[ T = \frac{(D-1)}{8 \pi \ell} \, M^{\frac{2}{D-1}} > 0 \quad \mbox{and} \quad
 S_{W} = 4 \pi \, \Omega_{D-2} \, M^{\frac{2(D-2)}{D-1}} > 0 \,. \] 
It is obvious from these, and it can also be seen from (\ref{ener31}) and (\ref{ent31}) for generic values of $z>2-D$, 
that the first law of black hole thermodynamics in the form \( dE = T dS_{W} \) does not hold for a generic member of 
this class of Lifshitz black holes. 

The failure of \( dE = T dS_{W} \) may stem from a number of reasons: 

i) It may be that the generalized Killing charge definition that follows from the ADT procedure may not at all be 
suitable for studying the thermodynamics of Lifshitz black holes. Obviously there are various other definitions of 
gravitational charges and specifically gravitational energy. One such definition relies on holographic renormalization 
technology that has been further advanced by techniques developed through the celebrated AdS/CFT duality for 
gravitational models that admit asymptotically locally AdS spacetimes \cite{ske}. It would certainly be of interest to 
employ such energy definitions on the solutions discussed here and to compare the results with what we have obtained 
above. If the first law is to hold in the form  \( dQ_{E} = T dS_{W} \), then such an energy calculation should give
\[ Q_{E} = \frac{(2-D)}{2 \ell} \, \Omega_{D-2} \, \frac{(3 z^2 + (D^2 - 4)) \, (D^2 - (2-3z)^2)}{q(z)} \, M^{2} \]
from (\ref{ent31}). Note in this case that the ``energy'' $Q_{E}$ is not necessarily positive definite either.

ii) The Wald entropy definition we have used here may not be quite correct after all. A strong indicator towards
this conclusion follows from the following observation: As can be seen by a careful scrutiny of $S_{W}$ in (\ref{ent31}), 
and also in the numerical calculation that follows (see also the example we consider in subsection \ref{sec42} below), it 
is obvious that the Wald entropy does not always satisfy the second law of black hole thermodynamics, i.e. that 
$S_{W} \geq 0$ does not always hold. Needless to say, this is quite contrary to what one expects from a correct definition 
of entropy.

iii) It is well known that in a generic quadratic curvature gravity theory in $D \geq 5$, there are also active
massive tensor and/or massive scalar gravitons apart from the usual massless gravitons. Even though it is
not apparent in the metric (\ref{lifbh}) (with (\ref{fofsec31})) itself with only the parameter $M$ to play with,
it may be that these active massive modes may modify the first law of black hole thermodynamics to a form
such as \( dQ_{E} = T dS + \Phi dq \), where $q$ is a scalar charge and $\Phi$ indicates its corresponding
potential.
\begin{table}
    \begin{tabular}{ | c | c | c | c | c | c | c | }    \hline
    $D$ & $z$ & $e_{1}$ & $t_{1}$ & $t_{2}$ & $s_{1}$ & $s_{2}$ \\ \hline
    5 & 2.05977 & 0.215524 & 5.05977 & 0.814175 & 0.143683 & 1.18582 \\ 
    6 & 2.509 & 0.11298 & 6.509 & 0.770932 & 0.0564899 & 1.22907 \\ 
    7 & 3 & 0 & 8 & 3/4 & 0 & 5/4 \\ 
    8 & 3.51981 & -0.113419 & 9.51981 & 0.739471 & -0.0378064 & 1.26053 \\ 
    9 & 4.05917 & -0.223906 & 11.0592 & 0.734083 & -0.0639732 & 1.26592 \\ 
    10 & 4.61188 & -0.330771 & 12.6119 & 0.731355 & -0.0826929 & 1.26864 \\ 
    11 & 5.17385 & -0.434265 & 14.1738 & 0.730056 & -0.0965034 & 1.26994 \\ \hline
    \end{tabular} \\
    \caption{The $z$ values for which \( dE = T dS \) and the corresponding coefficients in $E$, $T$ and $S$.}
\end{table}

As a final remark, we also want to point to the following argument: If one naively demands that 
\( dE = T dS_{W} \) does hold, then, given the spacetime dimension $D \geq 5$, one ends up with a 
polynomial of degree 5
\[ 2 (z-1) \, p(z) = (2-D) \, (z+D-2) \, (3 z^2 + (D^2 - 4)) \, (D^2 - (2-3z)^2) \]
in $z>2-D$, which is not easy to solve analytically. Nevertheless, it is possible to obtain
a numerical solution yielding real $z$ for spacetime dimension $5 \leq D \leq 11$.
This is given in Table I, where for convenience the coefficients \( e_{1}, t_{1}, t_{2}, s_{1}, s_{2} \)
have been utilized such that the relevant thermodynamical quantities read
\[ E = e_{1} \, \frac{M^2}{\ell} \, \Omega_{D-2} \,, \quad T = \frac{t_{1}}{8 \pi \ell} \, M^{t_{2}}
   \quad \mbox{and} \quad S_{W} = 4 \pi \, s_{1} \, \Omega_{D-2} \, M^{s_{2}} \,. \]
It is clear from Table I that the physically acceptable solutions, with $S_{W} \geq 0$, for which \( dE = T dS_{W} \) 
live in  $D=5$, $D=6$ and $D=7$ only. Here it should also be noted that the special case $D = 7$ and 
$z = 3$ is an exact analytic solution, with 
\begin{eqnarray*}
 \kappa & = & 1 \,, \quad \Lambda_{0} = \frac{21}{2 \ell^{2}} \,, \quad \alpha = - \frac{\ell^2}{160} \,, 
 \quad \beta = \frac{7 \ell^2}{160} \,, \quad \gamma = \frac{\ell^2}{32} \,, \\ 
 f(r) & = & 1 - \frac{M \ell^{4}}{r^{4}} \,, \quad E = S_{W} = 0 \quad \mbox{and} \quad 
 T = \frac{1}{\pi \ell} \, M^{3/4} > 0  
\end{eqnarray*}
precisely. We again expect this special solution to be a critical point of the relevant gravitational 
theory \cite{dllpst}.

\subsection{\label{sec32} The energy and the entropy of Lifshitz black holes for $z>1$}

This family of solutions can be found in section 3.2 of \cite{giri2}. They are again in the form (\ref{lifbh}), 
where the function $f(r)$ is
\begin{equation}
 f(r) = 1 - \frac{M \ell^{2(z-1)}}{r^{2(z-1)}} \,,
\label{fofsec32}
\end{equation}
with $M$ again as a free parameter. Similar to the class of solutions described in subsection \ref{sec31}, this
family is only defined for $D \geq 5$, but now the dynamical exponent has to satisfy $z > 1$. One again
has (\ref{param}) in terms of the parameters $\lambda$, $\beta_{1}$, $\beta_{2}$ and $\beta_{3}$, but now 
the latter are given by Eqs. (3.8b)-(3.8e) of \cite{giri2}.

We again simply state the result of a rather complicated energy calculation. One finds
\[ E = \left\{
\begin{array}{cl}
\infty \,, & 1 < z < \frac{D+2}{3} \\
\frac{(7-D) \, M^2}{6 \, \ell} \, \Omega_{D-2} \,, & z = \frac{D+2}{3} \\
0 \,, & \frac{D+2}{3} < z
\end{array} \right. \, . \]
Note that $E = 0$ for $D = 7$ when $z \geq 3$, and $E < 0$ for $D > 7$ when $z = (D+2)/3$.
As for the temperature and the entropy, one gets
\[ T = \frac{(z-1)}{2 \pi \ell} \, M^{\frac{z}{2(z-1)}} > 0 \quad \mbox{and} \quad
 S_{W} = 0 \quad \mbox{for all} \; D \geq 5 \,, \]
which implies that the first law of black hole thermodynamics in the form \( dE = T dS_{W} \) is plausible
for this class only for \( z > (D+2)/3 \). Note also that the solutions with $E = 0$ are possible critical points of 
the relevant theory \cite{dllpst}.

\subsection{\label{sec33} The energy and the entropy of Lifshitz black holes for $z<0$}

This family of solutions is given in section 3.3 of \cite{giri2}. They are defined only for a negative dynamical
exponent, so $z < 0$, and again, similar to the families studied in subsections \ref{sec31} and 
\ref{sec32}, they live only for $D \geq 5$. This family is a solution of the quadratic curvature gravity theory 
(\ref{act}), where
\begin{eqnarray*}
\kappa & = & 1 \,, \quad \Lambda_{0} = - \frac{z \big( 2 z^2 + 4 (D-2) z + (D-2)(D-3)  \big)}{4 \ell^2 (2 z + D -2)} \,, \\
\alpha & = & - \frac{\ell^2}{4 z (2 z + D -2)} \,, \quad \beta = 0 \,, \quad
\gamma = \frac{\ell^2 \big( 6 z^2 + 4 (D-2) z + (D-1)(D-2) \big)}{4 (D-3)(D-4) z (2 z + D -2)} \,,
\end{eqnarray*}
in our conventions. The function $f(r)$ in (\ref{lifbh}) is given by
\begin{equation}
 f(r) = 1 - \frac{M r^{z}}{\ell^{z}} \,,
\label{fofsec33}
\end{equation}
where $M$ is a constant. In this case, the energy is found to be
\[ E = \left\{
\begin{array}{cl}
\infty \,, & \frac{2-D}{3} < z < 0 \\
\frac{(D+7) \, M^2}{6 \, \ell} \, \Omega_{D-2} \,, & z = \frac{2-D}{3} \\
0 \,, & z < \frac{2-D}{3}
\end{array} \right. \, . \]
One also finds
\[ T = - \frac{z}{4 \pi \ell M} > 0 \, \quad \mbox{and} \quad
 S_{W} = 0 \quad \mbox{for all} \; D \geq 5 \,, \]
and the first law of black hole thermodynamics in the form \( dE = T dS_{W} \) holds
for this class of solutions only when \( z < (2-D)/3 < 0 \). Once again we may think of the solutions with $E = 0$ as 
possible critical points of the relevant quadratic curvature gravity theory \cite{dllpst}.

\subsection{\label{sec42} The energy and the entropy of four dimensional $z=6$ Lifshitz black hole}

The final spacetime we want to consider in this section is the four dimensional $z=6$ Lifshitz black hole
given in section 4.2 of \cite{giri2}. This is a solution of the gravitational field equations obtained from
(\ref{act}), where
\[ \kappa = 1 \,, \quad \Lambda_{0} = \frac{51}{2 \ell^{2}} \,, \quad 
 \alpha = - \frac{9 \ell^2}{64} \,, \quad \beta = \frac{25 \ell^2}{64} \]
in our conventions. The function $f(r)$ in (\ref{lifbh}) reads
\begin{equation}
 f(r) = 1 - \frac{M \ell^{4}}{r^{4}} \,,
\label{fofsec42}
\end{equation}
where $M$ is a constant. The energy is simply
\[ E = \frac{103 M^2}{8 \ell} \, \Omega_{2} \]
for this solution. The temperature and the entropy of this black hole are found as
\( T = M^{3/2}/(\pi \ell) \) and \( S_{W} = - 15 \pi \Omega_{2} \sqrt{M} < 0 \), which is
clearly unphysical, and the first law of black hole thermodynamics in the form \( dE = T dS_{W} \) 
does not hold with these values.

\section{\label{enerNMG} The conserved charges of the exotic solutions of NMG}

We next turn our attention to some solutions of the three dimensional NMG of \cite{hohm,more}. Among 
these solutions, there are those ones that asymptotically approach to the anti-de Sitter (AdS) space, 
e.g. the BTZ black hole \cite{btz} and the solutions presented in \cite{more,ott}. Since our charge 
expression (\ref{genchar}) already reduces to the original ADT formula \cite{des1,des2,des3} when 
the background is a space of constant curvature, it is simply unnecessary to calculate the conserved 
charges of these spacetimes here as this will only reproduce the results of \cite{clem,kore,ott,gott} 
in which these quantities were already computed using the ADT formula.

The first example, which is not asymptotically AdS, is the three dimensional $z=3$ Lifshitz black hole 
\cite{giri1} which is a cousin of the solutions considered in section \ref{enerLBH}. The energy of this 
solution has already been calculated in \cite{biz} using (\ref{ener}) (see \cite{biz} for details)\footnote{The 
energy was also computed in \cite{tonni} using the boundary stress tensor method. Moreover, the
thermodynamics of this solution was also studied in \cite{ykore} after a dimensional reduction to two
dimensional dilaton gravity.}. In terms of the conventions introduced in section \ref{enerLBH}, it is 
simply\footnote{In \cite{biz}, the parameters $\Omega_{1}$ and $\kappa$ were taken as 
\( \Omega_{1} = 2 \pi \ell \) and \( \kappa = 16 \pi G \) in accordance with the conventions of \cite{giri1}.}
\[ E = \frac{7 M^2}{\ell \kappa} \, \Omega_{1} \,, \]
whereas a careful calculation gives
\[ T = \frac{M^{3/2}}{2 \pi \ell} \quad \mbox{and} \quad S_{W} = \frac{16 \pi}{\kappa} \, \Omega_{1} \, \sqrt{M}, \]
which clearly shows that the first law of black hole thermodynamics in the form \( dE = (T/2) dS_{W} \) for
$D=3$ \cite{clem}, does not hold for the three dimensional Lifshitz black hole.

The next example that is of interest is the warped AdS$_3$ black hole solution \cite{clem}. This is
a solution to the NMG theory, i.e. the gravitational action (\ref{act}), in which
\[ D=3 \,, \quad \kappa = 8 \pi G \,, \quad \gamma = 0 \,, \quad \beta = - \frac{1}{m^2 \kappa} \,, \quad
 \alpha = - \frac{3}{8} \beta = \frac{3}{8 m^2 \kappa} \]
in our conventions. The metric of this black hole reads
\begin{equation}
 ds^2 = (1 - p^2) dt^2 - 2 (r + \omega (1 - p^2)) dt d\phi 
 + \Big[ (r + \omega)^2 + p^2 \Big( \frac{q^2}{1 - p^2} - \omega^2 \Big) \Big] d\phi^2 
 + \frac{dr^2}{p^2 (r^2 - q^2) \zeta^2} \,,
\label{wAdSBH}
\end{equation}
where the constants $\zeta$ and $p$ are given by
\[ p^2 = \frac{9 m^2 + 21 \Lambda_{0} - 2 m \sqrt{3 (5 m^2 - 7 \Lambda_{0})}}{4 (m^2 + \Lambda_{0})} 
\quad \mbox{and} \quad \zeta^2 = \frac{8 m^2}{21 - 4 p^2} \]
in terms of the bare cosmological constant $\Lambda_{0}$ and the NMG parameter $m^2 > 0$. As explained
in \cite{clem}, the metric (\ref{wAdSBH}) represents a causally regular warped AdS$_3$ black hole, provided 
\[ 0 < p^2 < 1 \quad \mbox{and} \quad \frac{35 m^2}{289} \geq \Lambda_{0} \geq - \frac{m^2}{21} \,, \]
which has a horizon located at \( r_{+} = q > 0 \).

The background metric relevant for our purposes can be chosen by setting $q \to 0$ and $\omega \to 0$ in
(\ref{wAdSBH}), which simply yields the metric of warped AdS$_3$ spacetime
\begin{equation}
 ds^2 = (1 - p^2) dt^2 - 2 r  dt d\phi + r^2 d\phi^2 + \frac{dr^2}{p^2 r^2 \zeta^2} \,,
\label{warpback}
\end{equation}
with the warping parameter $p$. The background metric (\ref{warpback}) clearly possesses two 
Killing isometries generated by the timelike Killing vector \( \bar{\xi}^{a} = - (\partial/\partial t)^{a} \), 
which can be employed for the computation of the energy of (\ref{wAdSBH}), and the spacelike
Killing vector \( \bar{\varsigma}^{a} = (\partial/\partial \phi)^{a} \), which should lead to the 
angular momentum of (\ref{wAdSBH}). One again chooses $\Sigma$ in (\ref{genchar}) to be 
a hypersurface of constant time, but one needs special care in determining the unit normal $n^{a}$
 of $\Sigma$. A diligent consideration yields 
\[ n^{a} = \frac{1}{p} \Big( \delta^{a}_{t} + \frac{1}{r} \delta^{a}_{\phi} \Big) \quad \mbox{or} \quad 
  n_{a} = - p \, \delta^{t}_{a} \,. \] 
Meanwhile, the choice of the unit normal $s^{a}$ of $\partial \Sigma$, which is located at $r \to \infty$, is
straightforward:
\[ s^{a} = p r \zeta \, \delta^{a}_{r} \quad \mbox{or} \quad s_{a} = \frac{1}{p r \zeta} \, \delta^{r}_{a} \,. \]
Finally, the measure of the one dimensional integral on $\partial \Sigma$ is simply 
\( \sqrt{|\sigma^{(\partial \Sigma)}|} = r \), all of which can also be deduced by putting the
background metric (\ref{warpback}) in the standard ADM form. Thus one has
\[ E \equiv Q(\xi) = - \frac{1}{\zeta} \int d\phi \, {\mathcal F}^{tr}(\xi) 
 = - \frac{\Omega_{1}}{\zeta} \Big( \lim_{r \to \infty} {\mathcal F}^{tr}(\xi) \Big) 
 \; \mbox{and similarly} \;
 J \equiv Q(\varsigma) = - \frac{\Omega_{1}}{\zeta} \Big( \lim_{r \to \infty} {\mathcal F}^{tr}(\varsigma) \Big) \,, \]
for the energy $E$ and the angular momentum $J$ of (\ref{wAdSBH}). Here, as before, $\Omega_{1}$ 
denotes the value of the one dimensional integration over the range of the angular variable $\phi$.

After a rather long calculation, we find
\[ E = \frac{16 p^2 (1 - p^2) \zeta \Omega_{1}}{(21 - 4 p^2) \kappa} \omega 
 = \frac{4 p^2 (1 - p^2) \zeta}{G (21 - 4 p^2)} \omega \,, \]
where special attention has been paid to using the `wrong-sign' (in the sense of \cite{more}) Einstein-Hilbert 
term in the action (\ref{act}) which is a requirement of NMG and we have set $\Omega_{1} = 2 \pi$. This result 
is identical to the one obtained in \cite{clem}, which was derived by first finding the angular momentum $J$ 
through other means and then imposing that the first law of thermodynamics in the form 
\( dE = T dS_{W} + \Omega dJ \) holds. 

The calculation of the angular momentum follows along the same lines and we find
\[ J = - \frac{\zeta}{4 G (21 - 4 p^2)} \Big( \frac{8 p^2 q^2}{1 - p^2} 
 + \frac{(1 - p^2)}{2 p^2} (21 - 29 p^2 + 24 p^4) \omega^2 \Big) \,. \]
This result is similar to but different than the one found in \cite{clem}: The first part that goes with $q^{2}$
is, but the second piece proportional to $\omega^2$ is not, equal to their counterparts given in \cite{clem}.
Having checked our calculation many times, we find this result quite perplexing and cannot account for
the discrepancy between it and that of \cite{clem}.

\section{\label{conc} Summary and Conclusions}

In this work we have calculated the energies of analytic Lifshitz black holes in higher dimensions. We have shown
that the first law of black hole thermodynamics does not always follow with the physical quantities we have found. 
In subsection \ref{sec31}, we have discussed in detail the possible reasons for this. We have also found the Lifshitz 
black holes with $E = 0$, which are possible critical points of the gravitational models under consideration. Separately, 
we have also computed the energy and the angular momentum of the warped AdS$_3$ black hole of the NMG theory. 
The energy we have found for the latter is in agreement with the one in \cite{clem}, but our expression for angular 
momentum is different than what \cite{clem} has.

Our focus in this work has mainly been on the Killing charge definition we introduced earlier \cite{biz}, and how
it can be used together with the first law of black hole thermodynamics. It would be interesting to use other energy 
definitions on the family of solutions we have considered and to compare the results with what we have found here. 
Perhaps separately from this, a more general issue, which is at least mystifying for us, is to understand how these 
different notions of gravitational energy definitions are related, if at all, to each other.

It would certainly be of interest if the rotating versions of these Lifshitz black holes could be found. Such
solutions would provide extra nontrivial examples in which our generalized Killing charge definition could
be employed. Such solutions would also provide for a wider playing ground on which the thermodynamics of
black holes of quadratic curvature gravity theories could be studied. We hope to return to this problem in
the near future.

\begin{acknowledgments}
This work is partially supported by the Scientific and Technological Research Council of Turkey (T{\"U}B\.{I}TAK).
\end{acknowledgments}

\end{document}